\documentclass[runinaddress,showpacs,aps,prb,superscriptaddress,a4paper,reprint,floatfix]{revtex4-1}

\usepackage{amsfonts}
\usepackage{amsmath}
\usepackage{amssymb}
\usepackage{bm}
\usepackage{color}
\usepackage{color,graphicx}
\usepackage{dcolumn}	
\usepackage{gensymb}
\usepackage{graphicx}	
\usepackage{latexsym}
\usepackage{mathcomp}
\usepackage{textcomp}
\usepackage{verbatim}
\usepackage{dcolumn}
\usepackage{ulem}

\providecommand{\STO}{SrTiO$_{3}$}
\providecommand{\tio}{TiO$_6$~}
\providecommand{\pis}{\textbf{P1}}
\providecommand{\pisk}{\textbf{P2}}

\definecolor{red}{rgb}{0.8,0,0.2}

\def\beeq{\begin{equation}}
\def\eneq{\end{equation}}
\def\beeqa{\begin{eqnarray}}
\def\eneqa{\end{eqnarray}}
\def\Ang{\AA}

\definecolor{adobe}{rgb}{.8,.6,.5}
\definecolor{mygreen}{rgb}{0.0,0.6,0.0}
\definecolor{blue2}{rgb}{0.0,0.0,0.8}
\definecolor{brown}{rgb}{0.6,0.3,0.0}
\definecolor{forest}{rgb}{0.0,0.4,0.0}
\definecolor{grass}{rgb}{0.0,0.55,0.25}
\definecolor{grass2}{rgb}{0.0,0.6,0.25}
\definecolor{gray}{rgb}{0.4,0.4,0.4}
\definecolor{grayish}{rgb}{0.2,0.2,0.4}
\definecolor{khaki}{rgb}{0.9,0.9,0.7}
\definecolor{lightteal}{rgb}{0.0,0.6,0.6}
\definecolor{lightyellow}{rgb}{1.0,1.0,0.5}
\definecolor{maroon}{rgb}{0.7,0.1,0.2}
\definecolor{navy}{rgb}{0.0,0.1,0.7}
\definecolor{olive}{rgb}{0.4,0.4,0.0}
\definecolor{orange}{rgb}{0.9,0.45,0.0}
\definecolor{peach}{rgb}{1.0,.8,.7}
\definecolor{purple}{rgb}{0.4,0,0.55}
\definecolor{teal}{rgb}{0.0,0.5,0.4}
\definecolor{turq}{rgb}{0.3,0.6,0.9}
\definecolor{violet}{rgb}{0.75,0,0.75}

\begin{document}
\DeclareGraphicsExtensions{.ps,.pdf,.eps}

\preprint{UPDATED: {\color{maroon} \today}}


\title{Accurate Modeling of the Cubic and Antiferrodistortive Phases of SrTiO$_3$ with Screened Hybrid Density Functional Theory}

\author{\firstname{Fadwa} \surname{El-Mellouhi}}
   \email{fadwa.el\_mellouhi@qatar.tamu.edu}
     \affiliation{Science Program, Texas A\&M at Qatar, Texas A\&M Engineering Building, Education City, Doha, Qatar}

\author{\firstname{Edward N.} \surname{Brothers}}
   \email{ed.brothers@qatar.tamu.edu}
     \affiliation{Science Program, Texas A\&M at Qatar, Texas A\&M Engineering Building, Education City, Doha, Qatar}

\author{\firstname{Melissa J.} \surname{Lucero}}
  \affiliation{Department of Chemistry, Rice University, Houston, Texas 77005-1892}

\author{Gustavo E. Scuseria}
  \affiliation{Department of Chemistry, Rice University, Houston, Texas 77005-1892}
  \affiliation{Department of Physics and Astronomy, Rice University, Houston, Texas 77005-1892}


\begin{abstract}

We have calculated the properties of \STO~(STO) using a wide array of density
functionals ranging from standard semi-local functionals to modern range-separated
hybrids, combined with several basis sets of varying size/quality.  We show how
these combination's predictive ability varies significantly, both for STO's
cubic and antiferrodistortive (AFD) phases, with the greatest variation in
functional/basis set efficacy  seen in modeling the AFD phase.  The screened
hybrid functionals we utilized  predict the structural properties of both phases
in very good agreement with experiment, especially if used with large (but still
computationally tractable) basis sets.  The most accurate results presented in
this study, namely those from HSE06/modified-def2-TZVP, stand as the most
accurate modeling of STO to date when compared to the literature; these results
agree well with experimental  structural and electronic properties as well as
providing insight into the band structure alteration during the phase
transition.  

\end{abstract}


\maketitle

\clearpage


\section{Introduction} \label{sec:intro} 

\begin{figure}
\includegraphics[width=3.1in, angle=0]{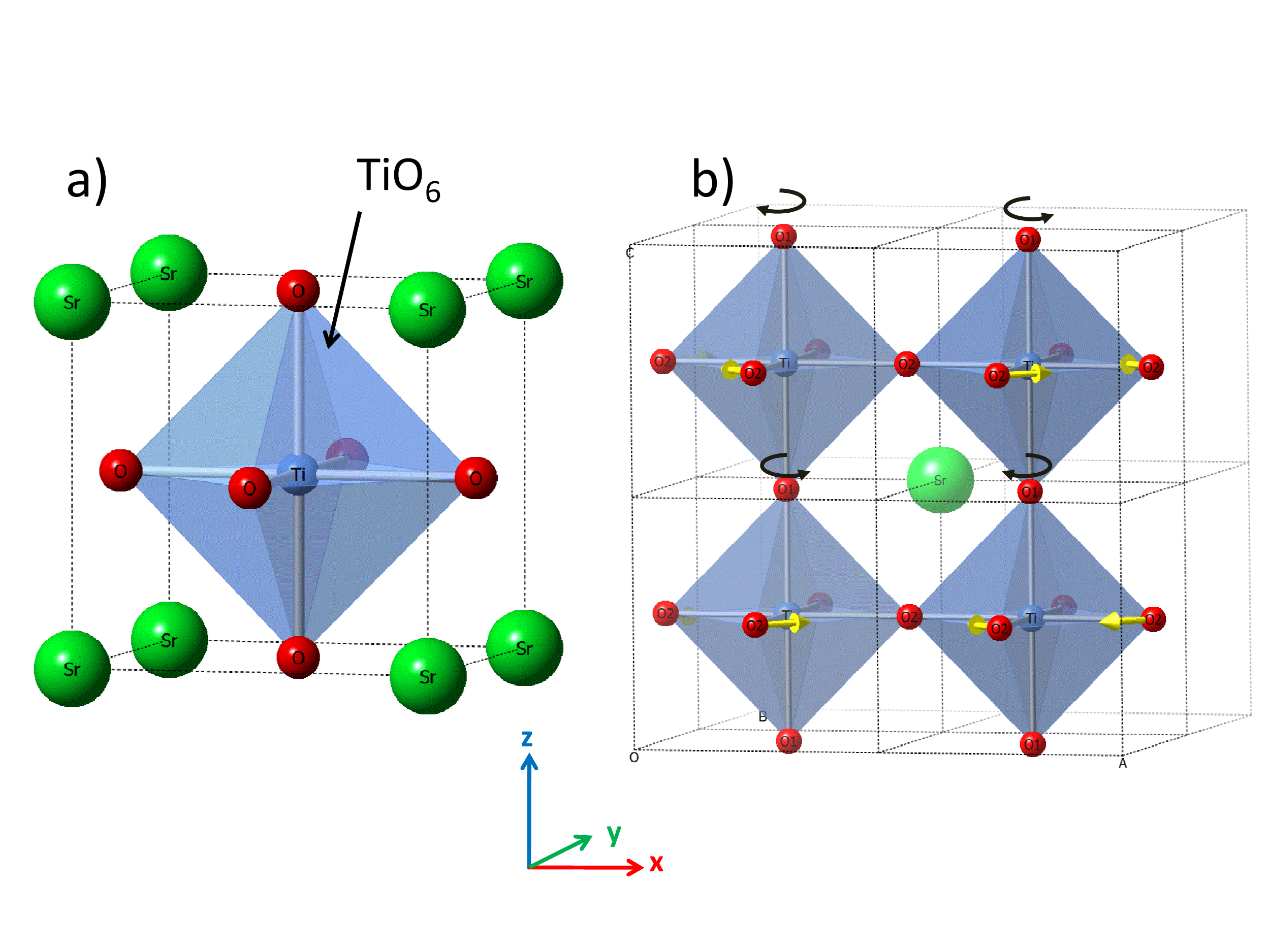} 
\caption{(Color online) \STO~unit cells for the a) cubic phase and b)
antiferrodistortive phase; b) shows the TiO$_6$ octahedra's rotation around the
[001] axis. Sr atom are in green, Ti are in blue and O are in red. The O$_1$ (equatorial) and
O$_2$ (axial) labels denote the non-rotating and rotating oxygens, respectively. }
 \label{fig:STOcell}
\end{figure}

Strontium titanate (\STO; STO) is a complex oxide perovskite of great
technological interest for its superconductivity,\cite{Ueno:2008} blue-light
emission,\cite{Kan:2005} photovoltaic effect,\cite{Zhou:2009} and so on.  Under
normal conditions, bulk \STO~crystallizes in a cubic perovskite structure; it
subsequently undergoes a second order phase transition at $T_c$=105 K to a
tetragonal structure with slightly rotated  oxygens around the z-axis, known as
the  antiferrodistortive~(AFD) phase (see Fig.~\ref{fig:STOcell}).  Many of the
interesting properties of STO, either in bulk or in  superlattices formed with
other metal oxides,  are  believed  to be caused by the  cubic to AFD phase
transition.  Examples of this attribution are STO's superlattice's high T$_c$
superconductivity~\cite{Reyren:2007, Caviglia:2008, Kozuka:2009} and its
colossal magnetoresistivity.~\cite{Gao:2009} First-principles calculations (see
Ref.~\onlinecite{Pentcheva:2010} and references therein) have indicated that the
strain-induced competition between octahedral rotation modes and the lattice
distortion in metal oxide superlattices are behind these interesting properties.
Thus, there is a considerable need~\cite{Borisevich:2010, Chang:2010} for
precise theoretical calculations of the structural and electronic properties of
complex oxides, as well as accurate estimation of the phase transition order
parameters, to understand and eventually exploit these phenomena.

The phase transition of STO is governed by two order parameters.  The primary
order parameter is the rotation angle of the TiO$_6$ octahedra ($\theta$).  The
experimentally measured~\cite{Unoki:1967} octahedral rotation of AFD STO is
1.4\degree~at 77 K and increases as the temperature drops toward the maximum
measured  value of 2.1\degree~at 4.2~K.  The octahedron's rotation is believed
to be almost complete~\cite{Jauch:1999tt} at around 50~K,  where
$\theta$=2.01$\pm$0.07\degree~was reported.\footnote{As the standard DFT
calculations done in this article not include temperature, we take the 0~K
experimental/target value to be 2.1\degree.}   The secondary order parameter is
the tetragonality of the unit cell ($c/a$), which increases from
1.00056~\cite{Cao:2000} to 1.0009~\cite{Heidemann:1973} as the temperature
decreases from 65~K to 10~K.\footnote{As temperature is not included in the
standard DFT work done here, we take the 0~K experimental/target value to be
1.0009.}   The  AFD phase can also appear in thin films of STO~\cite{He:2004ss,
He:2005ds, He:2003ig} at much higher T$_c$ than the bulk, depending  on the
substrate used,  the thickness of deposited STO film, the strain and  the
lattice mismatch.  For example, 10 nm of STO deposited on LaAlO$_3$ (LAO)
undergoes a transition to the AFD phase at \text{$T_c\cong$}~332~K.

As the simplest metal oxide perovskite,  STO has been extensively studied in the
last decades with different {\it ab initio} schemes.\cite{Zhuk09, Eglitis:2008,
Heifets:2006, Wahl08} However, it is still a challenging material for theory;
only a few of the  previously published works have been able to accurately
describe the structural and electronic properties of the both phases of STO.
The balance of this section will consist of a brief review of the theoretical
work performed to date.

Sai and Vanderbilt~\cite{Sai:2000hh} carried out one of the first LDA
calculations on STO using a plane-wave basis and ultra-soft pseudopotentials.
LDA predicted an exaggerated tetragonal AFD phase of STO, with octahedral
rotation angles of  6\degree,~significantly overestimating the
2.1\degree~rotation measured experimentally.~\cite{Unoki:1967}
Using LDA with other basis sets~\cite{Uchida:2003} shows similar issues,
predicting rotations up 8.4\degree. 

Wahl et al.~\cite{Wahl08} used  a plane-wave  basis while simulating STO with
LDA\cite{Vosko:1980}, PBE~\cite{Perdew:1996,Perdew:1997} and
PBEsol.\cite{Staroverov:2003, Staroverov:2004} (See  Section
\ref{sec:comp} for further descriptions of these density functionals).  LDA  underestimated experimental
lattice constants, while PBE overestimated them; both methods had band-gaps that
were seriously underestimated  compared to experiment.  This underestimation is
well known for these functionals; see {\it e.g.} Ref.~\onlinecite{Mori:2008} and
references therein.  PBEsol was found to reproduce accurately  the experimental
structure, but considerably underestimated the band gaps.  For the AFD phase,
the octahedral angle $\theta$ was found to be very sensitive to the functional
used; all three overestimate the AFD deformation, with LDA worse than PBE and
PBEsol splitting the difference.  Rondinelli et al.~\cite{Rondinelli:2010}
applied the LSDA+$U$ correction to cubic STO and found that while it corrects
the band gap, the calculated  octahedral rotation angle remains overestimated at
5.7\degree. To date, none of the post-DFT corrections which benefit
band-gaps have successfully  corrected the  octahedral rotation overestimation, and many authors attribute this  to the argument proposed  by Sai and Vanderbilt~\cite{Sai:2000hh} stating that  this 
  can be caused by the exchange and correlation terms  in  DFT  not capturing  quantum point fluctuations. 

Piskunov et al.~\cite{Pisk04} conducted one of the most complete and
comprehensive {\it ab initio} studies of STO, using Gaussian basis sets
specifically  optimized for modeling STO crystals.
This study of STO showed problems when modeling with pure DFT or pure HF, namely underestimated and overestimated band gaps, respectively; this is a well known problem.\cite{Janesko:2009} 
Hybrid functionals,
specifically B3PW\cite{Becke:1993} and B3LYP,\cite{Lee:1988} gave more
reasonable results, with direct band gaps overestimated by 5\% for B3PW and
3.5\% for B3LYP compared to experiment and indirect band gaps overestimated by
12\% for B3PW and 10\% for B3LYP.  (We will demonstrate  that an important part
of this overestimation can be attributed to the basis set employed; see
section~\ref{sec:basis}.) The hybrid functionals also gave the best agreement
with experiment for the lattice constant and the bulk modulus, and generally did
better than semilocal functionals in all categories.  This success of hybrid
functionals motivated more detailed calculations~\cite{Zhuk09,Eglitis:2008,
Heifets:2006} of the properties of the cubic and AFD phases of STO, again using
the optimized basis set of Piskunov et al.~\cite{Pisk04} and the B3PW
functional.  

Next, Wahl et al.~\cite{Wahl08} applied  the
Heyd-Scuseria-Ernzerhof~\cite{Heyd:2003, Heyd:2004} screened Coulomb hybrid
density functional (HSE) in a plane-wave basis set. HSE performed exceptionally
well, doing much better than any of semilocal functionals, as it gave a very
accurate estimate of both the structural and electronic  properties of the cubic
phase.  HSE also showed  excellent agreement with the experimental octahedral
angle and tetragonality of the unit cell which constitute, to our knowledge, the
most accurately computed  STO properties available in the literature for both
phases, prior to the current study.

 As noted above,
hybrid functionals have proved their effectiveness in  studying metal oxides,
but they are  computationally much more demanding than semilocal functionals.
While it would be  ideal to do  high accuracy {\it ab initio} 
on metal oxide superlattices  using complete basis sets and
large supercells, this is prohibitively expensive at the current level of
computer power. Screened hybrid functionals with only short range exact exchange
are computationally  less demanding; they allow the use of large supercells,
especially when used with localized basis sets such as Gaussian functions.  We
hope to use the most effective methods/basis sets from this study on more
complicated metal oxide systems, and thus we have concentrated on methods and
basis sets that would  be practical for those systems as well as the systems
currently under consideration.

This paper focuses on two tightly linked problems. We are interested in the
degree of completeness (or size) of the localized basis set necessary to
correctly simulate both phases of STO, and in the efficacy of recently developed
functionals (including  screened hybrids)  in  predicting  the properties of
STO.  To discuss these issues, the paper proceeds as follows: In Section
\ref{sec:comp}, we briefly describe the technical details before turning in
Section \ref{sec:basis} to the basis set optimization/modification technique we
used to make standard basis sets PBC-compatible.  In Section~\ref{sec:results},
we report the results of semilocal  and range separated hybrid functionals
applied to the cubic and the AFD phases of STO. We show also how the quality of
basis set affected the accurate prediction of the octahedral rotation angle in
the AFD phase  of STO.  Finally,  we  discuss the   results of our best
functional/basis set combination for STO, comparing them with previously
published theoretical and experimental data, with special emphasis on the effect
of varying the range separation parameter in the screened functionals.

%
\section{Computational Details} \label{sec:comp} 

All calculations were performed using a development version of the {\sc
Gaussian} suite of programs,\cite{gdv} with the periodic boundary condition
(PBC)\cite{Kudin:2000, Kudin:1998, Kudin:1998b} code used throughout.  A wide array of functionals were
applied, including: the Local  Spin Density Approximation~\cite{Vosko:1980}
(LSDA), the generalized gradient approximation (GGA) corrected functional of
Perdew, Burke and Ernzerhof~\cite{Perdew:1996,Perdew:1997} (PBE),  the
reparametrization of PBE for solids,
PBEsol,~\cite{Staroverov:2003,Staroverov:2004}  the revised meta-GGA of Tao,
Perdew, Staroverov and Scuseria\cite{Tao:2003,Perdew:2009} (revTPSS), and
finally a modern and highly parametrized meta-GGA functional,
M06L.~\cite{Zhao:2006,Zhao:2008} Two screened hybrid functionals were also
tested, namely the short-range exact exchange functional of  Heyd, Scuseria, and
Ernzerhof~\cite{Krukau:2006,Heyd:2003} (HSE, with the 2006 errata, also referred
to as HSE06) and  the exact exchange in middle-range functional\footnote{Originally introduced as HISS-B in Ref~\onlinecite{Henderson:2007} and called simply HISS as in Ref.~\onlinecite{Henderson:2008}} of Henderson,
Izmaylov, Scuseria, and Savin (HISS).\cite{Henderson:2007, Henderson:2008}
Because regular hybrids with unscreened exact exchange like B3LYP and B3PW  have higher computational cost compared to  screened
hybrids we decided to  exclude them from this test.

Gaussian basis sets of different quality have been tested for their ability to
simulate the properties of STO; the details of these tests and the modification
of the basis set was detailed enough to merit its own section,
section~\ref{sec:basis}. 

A few numerical considerations should be mentioned here.  During the initial (or
exploratory) calculations for the AFD phase, we found some dependence of octahedral rotation angle 
($\theta$) on initial atomic positions.  After further investigation, this can be attributed
to the geometric optimization convergence criteria.  Since $\theta$ is so small,
very stringent convergence criteria is required.\footnote{The standard RMS force
threshold in {\sc gaussian} for geometry optimizations is 450$\times 10^{-6}$
Hartrees/Bohr.  Using ``verytight'' convergence, this becomes 1$\times
10^{-6}$.} Another modified (versus the default) setting was that a  pruned
integration grid for DFT  of (99, 590) was employed, which corresponds to the
Gaussian option ``ultrafine''.  Note that this  grid is big enough for this
system to avoid any of the instabilities with M06L reported in the literature
with small grids.\cite{Johnson2009,Wheeler2010} To ensure this, we tested M06L
with a larger grid, without noticing any modification in  the calculated properties.  Thus
while ``ultrafine'' is sometimes insufficient for M06L, it is not for this
system.  Other numerical settings in {\sc gaussian} were left at the default
values, {\it e.g.} integral cut-offs, k-point meshes\footnote{Reciprocal space
integration used  12$\times$12$\times$12 {\it k }-point  mesh for the cubic unit
cell, while for the larger AFD supercell,  the default {\it k }-point  mesh of
8$\times$8$\times$6 was found to be sufficient.  }, SCF convergence
criterion,\footnote{Because we did geometry optimization, this was by default
set to ``tight'', or 10$^{-8}$.}  and the like.

Finally, the geometry of each phase is worth discussing briefly.  The starting
configuration for the cubic phase (see Figure~\ref{fig:STOcell}(a)) consisted of
the perovskite primitive cell containing 5 atoms at the experimental lattice
constant~\cite{Abra95} ($a_0$= 3.890 \AA).  For the AFD phase, we couldn't
simply use the 5 atom tetragonal unit cell with rotated oxygens and  the
lattice parameters set to $a=b\neq c$. A 20 atoms supercell/simulation cell  was
necessary,  as the phase transition requires a rotation of every pair of
neighboring TiO$_6$ octahedra in opposite directions
(figure~\ref{fig:STOcell}(b)). Thus, the volume of the AFD supercell  is about
four times the volume of the cubic phase  with tetragonal lattice constants
$a^*=b^*=\sqrt{2}a$ and $c^*=2c$, with $a$ and $c$ being the lattice parameters
of the 5 atoms tetragonal unit cell in the AFD phase. The
starting AFD structure  of STO   was taken from the experimental structure of
Jauch et al.~\cite{Jauch:1999tt} obtained  at 50~K and downloaded as CIF file
from the ICSD,\cite{ICSD}  with $a^*=b^*= 5.507$~\AA\ and $c^*=7.796$~\AA. The
starting rotation angle for TiO$_6$ octahedra was 2.1\degree\ while the $c/a-1
=10 \times 10^{-4}$.  Please note the  geometries were only starting points; as
mentioned above all geometries were optimized with the method/basis set under
consideration.  In order to avoid
introducing any errors coming from size effects or k-space integration, the
calculated properties of the AFD supercell are always compared with a 20 atoms
supercell constructed from  four  cubic primitive cells (without octahedral
rotation or tetragonality)  fully  relaxed using the same  {\it k }-point  mesh. It should be
noted that the supercell in the cubic phase is a local minimum and is  higher in energy than the supercell in the AFD phase for  all
reported calculations.
The final (reported) $\theta$  values were  determined from Ti-O$_2$-Ti angle
measurements, and any octahedral tilts can be estimated by measuring the
Ti-O$_1$-Ti angles (O$_n$'s subscript was defined in Figure~\ref{fig:STOcell}). Finally, all
geometric visualization was done using GaussView.~\cite{gv5}  

\section{Basis set efficiency for S\lowercase{r}T\lowercase{i}O$_3$} \label{sec:basis}

The challenge in selecting a basis set is always balancing accuracy with
computational cost.  
In molecular calculations, the computational cost of a gaussian basis set is determined by the number of functions used, while in PBC calculations the spatial
extent or diffuseness of the basis set also plays a major role.  The more diffuse a
basis set is, the larger the chunk of matter that must be included in the
calculations to avoid numerical issues.

Coupled with the
argument that the long density tail is more necessary for molecular work than
work in extended systems, it becomes obvious that basis sets developed for
non-periodic calculations can require modification for PBC use.  This  section
describes the basis set optimization/modification procedure we employed to find
the appropriate Gaussian basis sets to simulate periodic STO while keeping
within reasonable computational expense.  We based our evaluations of a basis
set's accuracy on  cubic STO results using the
Heyd-Scuseria-Ernzerhof~\cite{Heyd:2003, Heyd:2004} screened Coulomb hybrid
density functional  (HSE06).~\cite{Krukau:2006}

The obvious starting point was the basis sets used in previous
calculations/studies of bulk STO,  including:
\begin{itemize} 
    \item Gaussian-type basis sets  published by  Piskunov et
al.~\cite{Pisk00} in 2000, optimized  using the Hartree-Fock (HF) and 	density
functional theory (DFT)  with Hay-Wadt
pseudopotentials~\cite{HayW84a,HayW84b,HayW84c}  for Sr and Ti, denoted here as
{\bf P1}.\\
   \item The subsequently improved version of \pis~published by Piskunov et
al.~\cite{Pisk04}   in 2004, which expands of \pis~by adding polarization {\it
d}-functions to oxygen and making the Ti {\it s} and {\it p}  functions more
diffuse,  denoted here as \pisk. 
\end{itemize}

\begin{table}[!htb]
\caption{The electronic and structural properties of cubic  \STO~ computed with
HSE06~\cite{Krukau:2006} and different basis sets.  Please see the text for
basis set naming conventions.}
\begin{ruledtabular}
\begin{tabular}{lcccccl}
\label{tab:basis}
Basis set &P1   &P2 &SZVP &TZVP   &Experiment \\
\\
\hline
\\
Direct gap (eV)   &3.87   &3.80   &3.59   &3.59     &3.75\footnotemark[1]\\
Indirect gap (eV)   &3.53   &3.46   &3.18   &3.20     &3.25\footnotemark[1] \\
a$_0$($\text{\Ang}$)   &3.900   &3.908  &3.887   &3.902   &3.890\footnotemark[2], 3.900\footnotemark[3]\\
B(GPa)   &198   &194   &204   &193   &179\footnotemark[2],\\
&&&&&179$\pm{4.6}$\footnotemark[4]
\end{tabular}
\end{ruledtabular}
\footnotetext[1]{Reference~\onlinecite{Benthem:2001}.} \footnotetext[2]{Reference~\onlinecite{Abra95}.}
\footnotetext[3]{Reference~\onlinecite{Hell69}.} \footnotetext[4]{Reference~\onlinecite{Fischer:1993sd}.}
\end{table}
Tests  on \pis ~and \pisk~were done with HSE06, because it has been found to
give the best results versus  experiment for both structural and electronic
properties\cite{Wahl08} in older calculations.  Both \pis ~and \pisk
~reproduce the experimental equilibrium lattice constants~\cite{Abra95} (see
Table~\ref{tab:basis})  almost perfectly.   
Cubic  STO modeled with
\pis~ has a slightly  higher bulk modulus compared to \pisk,
although the difference between
the two basis sets in fairly minimal for structural properties.  
A more important effect is observed for
the electronic properties: \pis ~and \pisk ~overestimate the direct band gap
of STO by 0.12 and 0.05 eV respectively, and seriously overestimate the
indirect band gap by 0.28 and 0.21 eV. 

It is easy to see that the \pisk ~basis set employed with HSE06 lead to results
that are  closer to experiment than \pis, a fact noted by Piskunov\cite{Pisk04}
for a number of functionals. The more important point is that increasing the
size/quality of the basis set made a noticeable change in the results; the
immediate question is whether another increase in basis set size would bring about similar improvement.  In other words, using
polarization {\it d}-orbitals  for O and diffuse functions for Ti  improved the
HSE06 results, and imply that further improvement could potentially be achieved
if more basis set improvements are implements, {\it e.g.}  including titanium core
electrons and/or adding more diffuse functions for oxygen.

We decided to optimize some of the {\bf Def2-}~\cite{Weigend:2005} series of
Gaussian basis sets for use in bulk STO calculations.  The original {\bf Def2-}
basis sets for the atoms of interest in this project included small-exponent
diffuse functions ($\alpha_{min}$ less than 0.10) that are spatially quite
extended; as stated above, this long tail is necessary to improve the DFT
results for molecules but not necessary for crystals.~\cite{Heyd:2005, Strain:1996} Basis sets with large spatial
ranges dramatically slow down the calculation of Coulomb contributions to the
total energy of crystals.  Thus, to be useful in PBC calculations,  {\bf Def2-}
basis sets must be modified by removing some of the most diffuse functions. 

The series of {\bf Def2-} basis sets  are  available up to  quadruple zeta
valence quality for a large set of elements.\cite{Weigend:2005, Weigend:2006} In
the original optimizations, the oxygen, strontium and titanium basis sets were
optimized (using HF and DFT) versus the properties of SrO, TiO and  TiO$_2$ molecules.
Strontium has the inner shell electrons replaced with small core
pseudopotentials,~\cite{Kaupp:1991} while the other two atoms utilize all
electron basis sets; this differs from \pis~and \pisk~which uses
pseudopotentials on titanium as well.  In general, {\bf Def2-} basis sets are
larger and more expensive than \pis ~and \pisk~basis sets,  but are expected to
give a  better representation of both phases of STO due to greater
``completeness.''

To make a {\bf Def2-} basis set applicable to PBC, the first step is selecting a
maximum allowable diffuseness, or equivalently the smallest acceptable Gaussian
orbital exponent, $\alpha_{min}$.  The larger the value of $\alpha_{min}$, the
faster the calculations become, but if $\alpha_{min}$ is set too high,
significant degradation of physical property prediction results. After the threshold is defined, one
pass is made through the basis set to reset all $\alpha < \alpha_{min}$ to
$\alpha_{min}$, and then a second pass is made through the basis set to remove
any redundancies.  Note that after modifying or deleting an element of a
contracted basis set, we rely on the code's internal renormalization code, {\it
i.e.} no attempt is made to reoptimize contraction coefficients.  

We first began with the largest {\bf Def2-} basis sets, Def2-QZVP and
Def2-QZVPP, but these were found to be computationally intractable for bulk STO
even for $\alpha_{min}$ as big as 0.2, and previous experience have shown that
$\alpha_{min}$ larger than 0.2 causes physically unacceptable results.  We then
moved to the smaller basis sets, Def2-TZVP and Def2-SZVP.  We first set
$\alpha_{min}$=~0.12, but found this made the calculations very slow. Our tests
showed that $\alpha_{min}$=~0.15 constitutes a more computationally efficient
choice without important loss in accuracy. 

Henceforth,  the Def2-TZVP and Def2-SZVP, with $\alpha_{min}$ modified and 
redundant {\it s} functions removed,  will be denoted {\bf TZVP} and {\bf SZVP},
respectively.

Table~\ref{tab:basis}  summarizes the calculated electronic and structural
properties of cubic STO using our basis set modifications  as well as the
aforementioned \pis~and \pisk.  The optimized basis sets {\bf SZVP} and {\bf
TZVP} give an overall excellent agreement with experiment:~\cite{Abra95}  direct
band gaps are now underestimated by 0.16 eV  while  indirect band gaps are now
underestimated by $\sim$0.05 eV.   These two new basis sets are larger than the
previously utilized \pis ~and \pisk, are more accurate for indirect gaps, as
well for other measured properties, and due to their greater size are expected
to be closer to the upper limit of HSE06 accuracy for this system.  Note also
that the electronic properties of STO remain almost unchanged by moving from a
{\bf SZVP} to {\bf TZVP} basis set.  The deviation from the experimental lattice
constant do not exceed 0.07 and 0.3\% for {\bf SZVP} and {\bf TZVP} respectively
but is more substantial for the bulk modulus reaching 14\% for {\bf SZVP} and
8\% for {\bf TZVP}.  Finally, the same series of basis set optimizations were
also performed using HISS and M06L functionals, which lead to the same
conclusions regarding the basis set efficiency; these are not presented here for
space reasons.  

Before moving on to the results section, a brief mention of the expense of the
various basis sets should be included.  In term of relative CPU time, one SCF
cycle takes about  12  units for {\bf TZVP} compared to 6  units for {\bf SZVP}
and 1 unit for  \pisk.  All of these basis sets still have potential uses; {\bf
SZVP} or \pisk,  for example,  might be very useful for  a rapid  investigation
of the electronic properties of some complex STO systems. But, in term of
completeness, {\bf TZVP} is the most complete and the closest to the planewave
basis set limit, followed by {\bf SZVP}, then \pisk. 

\section{Results: Basis set and functional evaluation }
 \label{sec:results}

In this section we present the calculated properties of \STO, always
discussing the results of each functional using the {\bf TZVP} basis set first,
followed with a discussion of the sensitivity of the functionals to 
smaller basis sets, namely {\bf SZVP} and \pisk. 

\subsection{Structural  properties of cubic \STO} %
\label{sec:struct}
%
\begin{table*}[!htbp]
\caption{ Computed lattice parameter $a_0$(\AA) and bulk modulus B(GPa) for Cubic STO using different combinations of functionals  and basis sets compared to experiment. }
\begin{ruledtabular}
\begin{tabular}{lcccccccc}
\label{tab:ela-cub}

    &HSE06   &HISS   &M06L   &revTPSS  &LSDA        &PBE        &PBEsol     &Experiment\\

\\
\hline
\\
{\bf {a$_0$}(\AA)}                                                &&&&&&&&3.890\footnotemark[1], 3.900\footnotemark[2]\\
TZVP    &3.902   &3.883  &3.925  &3.921 &3.862  &3.941  &3.897  \\
SZVP    &3.887   & 3.869  &3.909  &3.903 &3.845  &3.924  &3.881  \\
P2  		&3.908  &3.891  &3.930   &3.920 &3.870       &3.946  &3.903     \\
\\
{\bf B(GPa)}                                                    &&&&&&&&179\footnotemark[1],\\
TZVP    &193         &206       &187        &180 &201       &169     &184    &179$\pm{4.6}$\footnotemark[3]    \\
SZVP    &204         &218       &198        &193 &214      &180       &196        \\
P2  		&194        &205       &191        &184 &203      &173       &187        \\
\end{tabular}
\end{ruledtabular}
 \footnotetext[1]{Reference~\onlinecite{Abra95}.}
\footnotetext[2]{Reference~\onlinecite{Hell69}.}
 \footnotetext[3]{Reference~\onlinecite{Fischer:1993sd}.}
\end{table*}
The calculated equilibrium lattice constants and the bulk moduli of cubic STO
using different functionals and basis sets are reported in
Table~\ref{tab:ela-cub}. Unless specified, the deviation of theory  from
experiment will be always referred to the data of Abramov et al,~\cite{Abra95} {\it i.e.} treated as the target value.
Focusing first on
the {\bf TZVP} results, we observe that the   screened hybrids HSE06 and   HISS give lattice parameters in excellent agreement with experiment. 

The calculated  bulk modulus using HSE06 is fairly
close to the  experimentally reported values, although overestimated  by 8\%.
(The same magnitude of overestimation  have been also reported in the HSE
planewave calculations of Wahl et al.~\cite{Wahl08}.) However, a larger bulk
modulus overestimation of 15\% is observed for  HISS, which constitutes the largest deviation from experiment among all
the studied functionals. 

M06L and revTPSS predict slightly higher equilibrium lattice constants than
screened hybrids do, but their bulk moduli are closer to experiment, with revTPSS being especially close.
  LSDA underestimates the lattice constant by 0.03 \AA, while PBE
predicts lattice constants 0.05 \AA~larger than experiment. PBEsol is in excellent agreement with the experimental lattice constant.
Thus PBEsol corrects the LSDA underestimation and the PBE overcorrection to LSDA
for lattice constants; in addition, the PBEsol  bulk modulus deviate by less
than 3\% from experiment, while LSDA and PBE are off by 11\% and 12\%,
respectively.  This is an example of PBEsol meeting its purpose, as it
improves the PBE lattice constant and bulk modulus for the cubic phase,
approaching very closely the experimental data. 

Turning now to the functional sensitivity to basis set size, we observe  from
the HSE06 results that {\bf SZVP} basis set predict bond lengths that are very slightly 
shorter than  the {\bf TZVP } and a bulk modulus that is ~6\% higher. As such,
{\bf SZVP} predicts \STO~to be 14\% harder than experiment.  \pisk~behave in the
opposite direction, predicting slightly longer bonds when compared to  {\bf
TZVP}, while the bulk moduli are only 1GPa higher. From table~\ref{tab:ela-cub},
this sensitivity of HSE06 to smaller basis set  can be generalized to M06L,
revTPSS and the semilocal functionals LSDA, PBE and PBEsol.  

Finally, it should be noted that  PBEsol results offer the best agreement with
experimental structural properties~\cite{Abra95} of \STO~among all the studied
functionals with the {\bf TZVP} basis set, followed by the screened hybrid
HSE06 and the meta-GGA  revTPSS. 

\subsection{Electronic properties of cubic \STO} 
\label{sec:elect_cub}
%
%

\begin{table*}[!htb]
\caption{Direct and indirect band gaps computed for Cubic \STO~using different basis sets and functionals compared to experiment. }
\begin{ruledtabular}
\begin{tabular}{lcccccccc}
\label{tab:elec_cub}

    &HSE06  &HISS   &M06L   &revTPSS &LSDA        &PBE        &PBEsol     &Experiment\\

\\
\hline
\\
{\bf Direct gap(eV)}                                           &&&&&&&&3.75~\cite{Benthem:2001}\\
TZVP    &3.59        	&4.39     &2.51       &2.24 &2.08       &2.11       &2.10        \\
SZVP    &3.59       	&4.45     &2.53       &2.28  &2.12       &2.14       &2.14       \\
P2  		&3.80         &4.56      &2.63   &2.52 &2.34       &2.33       &2.34          \\
\\
{\bf Indirect gap(eV)}                                                      &&&&&&&&3.25~\cite{Benthem:2001}\\
TZVP    &3.20          &3.98     &2.09       &1.87 &1.75       &1.74       &1.75       \\
SZVP    &3.18          &4.03     &2.10       &1.89  &1.76      &1.75       &1.76       \\
P2       &3.46           &4.22     &2.24       &2.17 &2.04       &1.99       &2.02       \\
\end{tabular}
\end{ruledtabular}
\end{table*}

The computed electronic properties of \STO~are summarized in
table~\ref{tab:elec_cub}.  As expected, HSE06 gives an excellent  estimate of the
electronic properties when used with the large  {\bf TZVP} basis sets.
Deviations from the experimental values are 0.15 eV for the direct gap and 0.05
for the indirect gap.  A cursory glance over the rest of
table~\ref{tab:elec_cub} indicates that no other functional was comparable
to HSE06's efficacy for band gaps, {\it i.e.} everything else we tried had much
larger errors.  

The midle range screened hybrid HISS tend to overestimate the
direct and indirect band gaps by  0.35 and 0.73 eV, respectively.  M06L and revTPSS tend to
underestimate  both band gaps,  by an average of $\sim$1.2 and $\sim$1.4 eV
respectively.   The semilocal functionals LSDA, PBEsol, PBE underestimate the
experimental  band gaps  by  an average of ~45\% or 1.5 eV.  This  was
expected, and is in agreement with the  behavior observed earlier in the
literature for this system.~\cite{Wahl08}  It can be easily seen from these
results that HSE06 is the best functional choice for investigating this system.

Turning to basis set sensitivity, it can be seen from the HSE06 numbers that
band gaps are nearly unaffected by using the smaller {\bf SZVP} basis sets, but
when used with the still smaller \pisk ~basis set, direct and indirect band gaps
increase by  $\sim$0.25 eV versus  {\bf TZVP}.  The predicted direct band gap
becomes closer to experiment when using \pisk~and HSE06, probably due to a
cancellation of errors effect, while the indirect band gap is noticeably worse.
This same sensitivity holds for almost every other functional, with {\bf SZVP} and
{\bf TZVP} giving about the same band gaps and \pisk~opening the band gaps up by
a few tenths of an eV.  M06L appears to be slightly less sensitive; no obvious
reason for this exists.

\subsection{Stability of the AFD phase of STO} \label{sec:AFD} 

This section examines the stability of the AFD phase of STO as calculated by the
various  functionals and basis sets previously tested for the cubic phase. The
functional/basis set combinations tested face the challenge of predicting the AFD octahedral rotation angle, $\theta$, as well as the tetragonality parameter
$c/a$, which as shown in section~\ref{sec:intro} is not trivial. The AFD phase
order parameters are evaluated from the relaxed 20-atoms AFD supercells as
described  in section~\ref{sec:comp}.  
The performance  of each functional with {\bf TZVP}, and then an analysis of the
functional's  sensitivity to the smaller basis sets is presented  in turn.

\begin{figure*}
        \includegraphics[width=3.1in, angle=0]{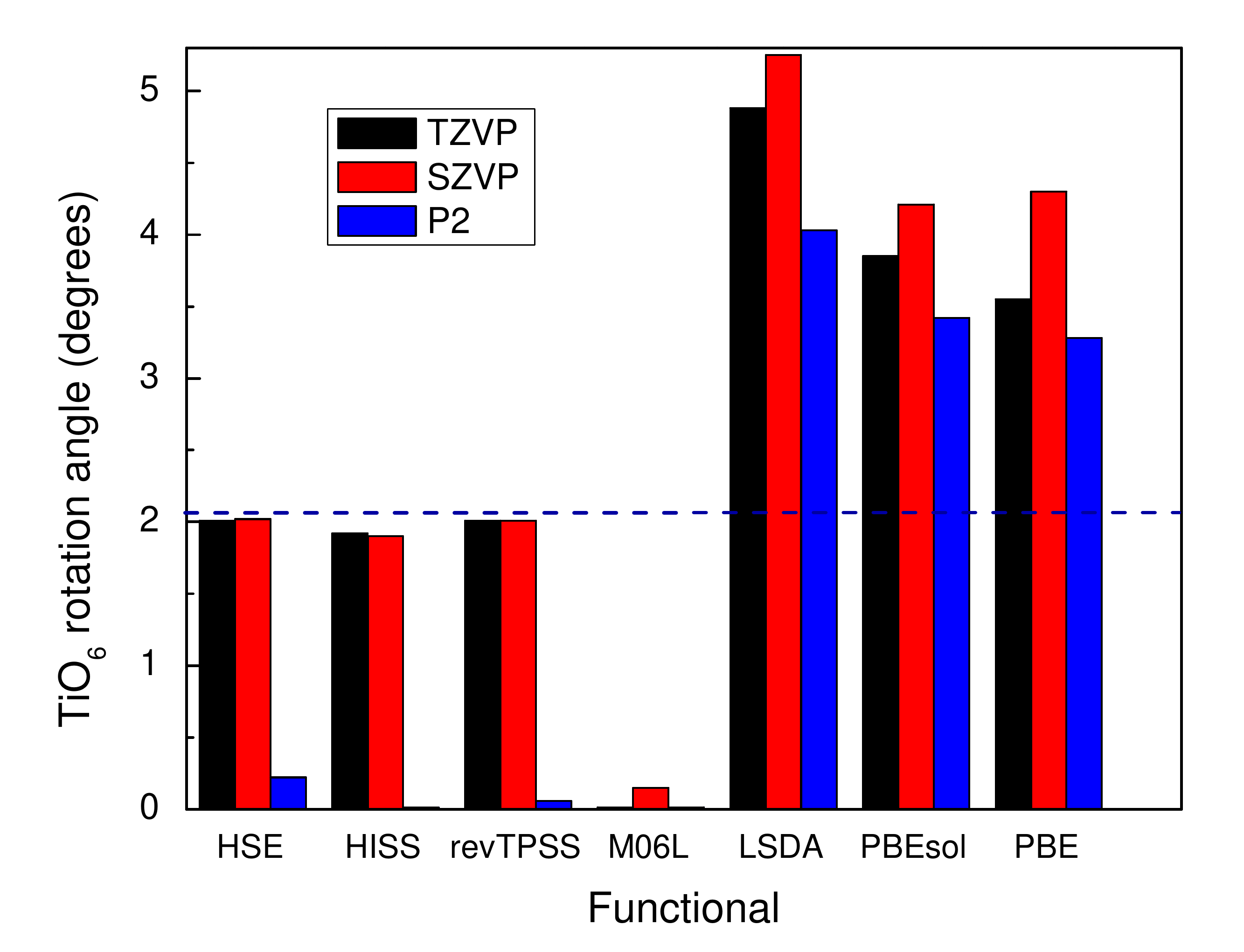} \includegraphics[width=3.1in, angle=0]{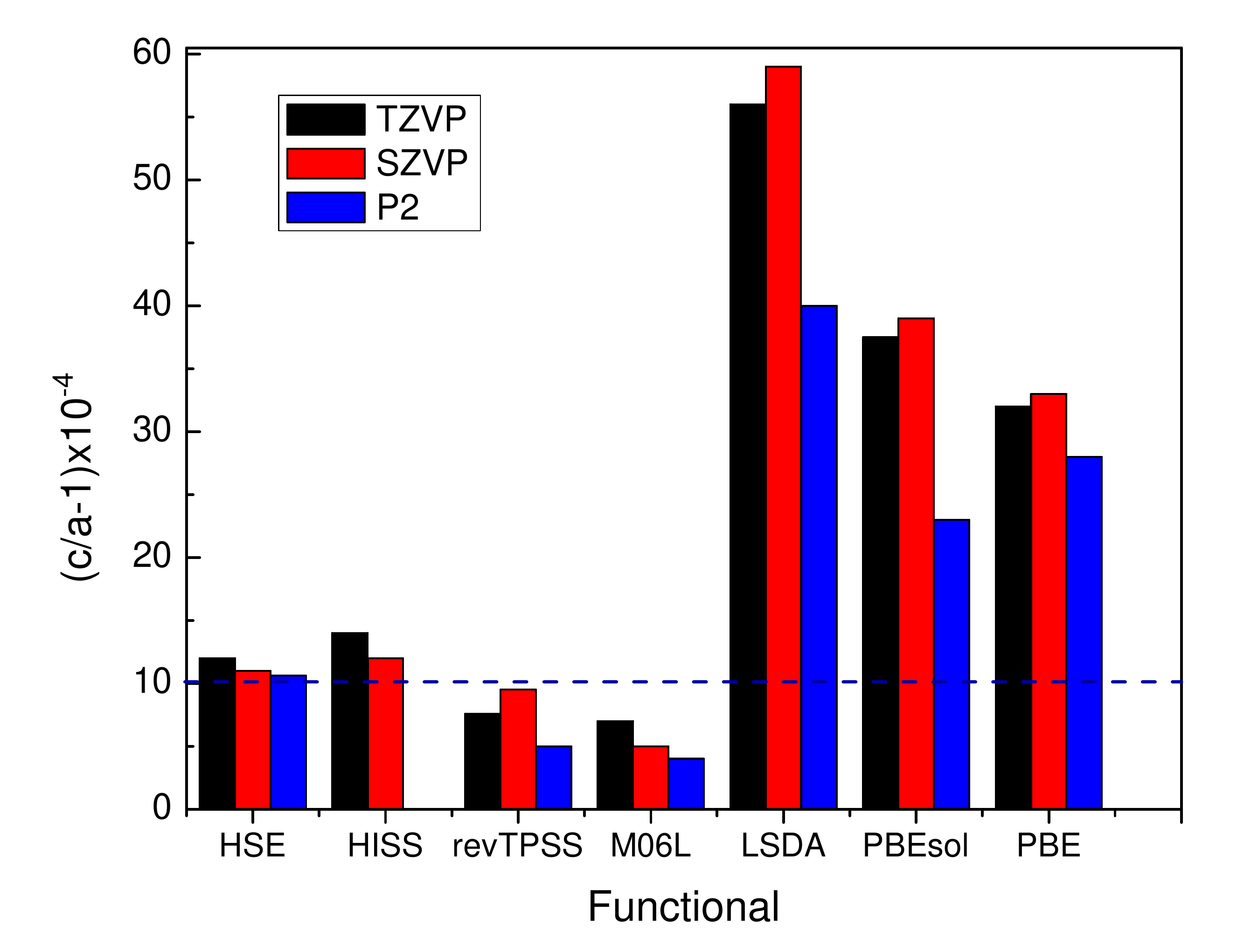}
       \caption{(Color online) Performance of different functional/basis set combinations  in predicting the order parameters of the AFD phase transition  in  STO.  Dashed  lines depict the experimental octahedral angle measured  at 4.2~K from Ref.~\onlinecite{Unoki:1967} (left) and the tetragonality parameter obtained at 50~K from Ref.~\onlinecite{Jauch:1999tt} (right).  }
 \label{fig:histo}
\end{figure*}

Figure~\ref{fig:histo} shows that the screened hybrid functional HSE06 is
excellent  for the structural properties of AFD, as it was for the cubic phase. Both the
rotation angle $\theta$ and the $c/a$ ratio are in very good agreement with
experiment.  These properties are not significantly affected when {\bf SZVP} is
used, but HSE06/\pisk~predicts a very very small angle for the AFD phase, while retaining a good $c/a$.  This is one area where {\bf TZVP} noticeably outperforms \pisk~ with HSE06.

HISS and revTPSS behave as HSE06 for both  {\bf TZVP} and {\bf SZVP}, giving a good
estimate of both order parameters. However,  they  demonstrate a higher
sensitivity to the smaller \pisk~basis set and required the use of  a very stringent convergence criterium to finally relax the structure back  to a pseudocubic  phase with  $\theta\approx$0.
On the other hand, M06L predicts the AFD phase to be unstable, and relaxes to a non-rotated structure  regardless of the basis set
used.

\begin{table*}[!htb]
\label{tab:AFD}
\caption{Structural and electronic properties  of the antiferrodistorsive phase of \STO~compared to previously simulated data and experiments.   $a^*$ and $c^*$ are the lattice parameters of the  AFD supercell and $c/a=c^*/{\sqrt{2}a^*}$. $\Delta E=E_{Cubic}-E_{AFD}$ represent the gain in total energy  after the cubic to AFD  phase transition, while $\Delta E_g$ denote the corresponding  increase in the band gap.}
\begin{ruledtabular}
\begin{tabular}{llllllllll}
      		&LSDA    &PBE &PBEsol  &HSE06   &HISS   &revTPSS &M06L&Experiment\\
\hline
\\
 $a^*$(\AA)                                        \\
Present 							&5.449	&5.568	&5.500	&5.515	&5.448	&5.543	&5.551	&5.507\footnotemark[5] \\
Ref.~\onlinecite{Wahl08}\footnotemark[1]				&5.440	&5.562	&5.495	&5.515	&	&	&	&	&\\
\\
$c^*$(\AA)                                        \\
Present 							&7.727	&7.900	&7.812	&7.809	&7.772	&7.846	&7.862	&7.796\footnotemark[5] \\
Ref~\onlinecite{Wahl08}\footnotemark[1]				&7.755	&7.897	&7.818	&7.808	&	&	&	&	&\\
\\
 $(c/a-1)\times10^{-4}$                                                      \\
Present 							&27	&32	&44	&12	&14	&7.6	&7	 &10\footnotemark[5] \\
Ref.~\onlinecite{Wahl08}\footnotemark[1]				&80	&40	&60	&10	&	&	&	&	&\\
Others                                          &40\footnotemark[4]\\
\\
 $\theta$(\degree)	\\
Present 					&4.14	&3.54	&3.81	&2.01	&1.92	&2.01	&0	&2.01$\pm$0.07\footnotemark[5] \\
Ref.~\onlinecite{Wahl08}\footnotemark[1]			&6.05	&4.74	&5.31	&2.63	&	&	&	&2.1\footnotemark[6] \\
Others                  &8.40\footnotemark[2], 6\footnotemark[3]\\
                            &4\footnotemark[4]\\
\\
 $\Delta E \times 10^{-5}$ (eV)                                                      \\
Present 			&1796	&854	&44		&35		 &578	&258 &122\\
Ref.~\onlinecite{Wahl08}\footnotemark[1]				&1900	&700	&1100	&200	&	&	&	& &\\
\\
Indirect band gap (eV)                                                    \\
Present 						&1.820	&1.787	&1.808	&3.227		& 3.995 &1.890	&2.060	&3.246~\cite{Yamada:2010}\\
Ref.~\onlinecite{Wahl08}\footnotemark[1]		&1.970	&1.790	&1.930	&3.110	&		& &	&3.160~\cite{Hasegawa:2000}\\
\\
 $\Delta E_g$ (meV)                                                     \\
Present 				&75	&49	&58	&27		&15 &15	&30	&50\footnotemark[7]\\
Ref.~\onlinecite{Wahl08}\footnotemark[1]		&160	&10	&110	&40	&	&	&	\\
\end{tabular}
\end{ruledtabular}
\footnotetext[1]{plane-wave calculation using a different HSE screening parameter.}
\footnotetext[2]{Reference~\onlinecite{Uchida:2003}.}
\footnotetext[3]{Reference~\onlinecite{Sai:2000hh}.}	
\footnotetext[4]{Reference~\onlinecite{Hong:2010} using numerical atomic orbitals.}
\footnotetext[5]{Reference~\onlinecite{Jauch:1999tt} (at 50K).}
\footnotetext[6]{Reference~\onlinecite{Unoki:1967} (at 4.2K).}	
\footnotetext[7]{Reference~\onlinecite{Yamada:2010} difference between 85~K and 8~K measured gaps.}
\end{table*}

The semilocal functionals LSDA, PBEsol and PBE all overestimate the tetragonality
of the AFD phase by  predicting  $\theta$  and $c/a$ almost twice the size of the 
experimental results.  The highest overestimation was observed for LSDA,
followed by PBEsol then PBE. Note that our result here is in excellent qualitative
agreement with the behavior found in the  planewave calculations of Wahl et
al;~\cite{Wahl08}  quantitatively, however, the LSDA, PBEsol and PBE octahedral angles with
{\bf TZVP} are 25-30\% lower than the planewave  results~\cite{Wahl08,
Sai:2000hh, Uchida:2003} (for a detailed numerical comparison see
table~\ref{tab:AFD},  and ref.~\onlinecite{Wahl08} has additional
comparison with experiment).  Similar behavior has been recently
published~\cite{Hong:2010} for LSDA calculation with  finite-range numerical
atomic orbitals  using a double-$\zeta$ polarized basis set. 
This indicates that  localized basis sets tend to reduce the  AFD octahedral rotation compared to plane-waves but do not succeed to suppress the DFT overestimation.

 When used with the {\bf SZVP} basis sets, the LSDA, PBE and PBEsol rotation angles are larger than the TZVP ones.
Furthermore, when  LSDA, PBEsol and PBE  are used with the \pisk~basis set,  we
observe a small and coherent reduction in the octahedral rotation angle of the
AFD structure compared to {\bf TZVP} results.  This demonstrates  that semilocal
functionals have different  degrees of sensitivity to the quality of the
localized  basis sets used, but the error caused by functional choice is always the more important source of error.
Thus the functionals examined here will lead to exaggerated AFD $\theta$ values for  all
basis sets considered. 


\section{Discussion: Physical properties of STO}\label{sec:disc} 

Before talking about specific issues, there are a few general conclusions we
can reach from examining the results in section~\ref{sec:results}

\begin{enumerate} 
  \item HSE06/\pisk~did a good job in describing accurately the structural
properties for the cubic phase as well as providing a descent estimation of the
band gap.  However, the failure of HSE06/\pisk~to correctly model the structure of the AFD
phase indicates  that it must be abandoned as a useful combination for
this and related systems.  
  \item HSE06/{\bf SZVP} has the drawback of predicting a stiffer \STO~in the
cubic phase, although it predicts electronic properties as well as {\bf TVZP}.  It
also predicts a stiffer AFD structure, but the octahedral angle and $c/a$
parameters are very good.

  \item   HSE06/{\bf TZVP} gave the best agreement with  experiment for the
cubic phase and for the AFD phase.  It is definitely the most reliable
combination of functional and basis set among all studied variations.  Thus
HSE06/{\bf TZVP} can be used with confidence on more complicated structures, as
well as to understand the change in the electronic structure during the cubic to
AFD transition for this system.  More concisely, we believe that this combination is an accurate
enough functional in a good enough basis set to explain phenomena in metal
oxides.
\end {enumerate}

\begin{figure}
        \includegraphics[width=3.0in, angle=0]{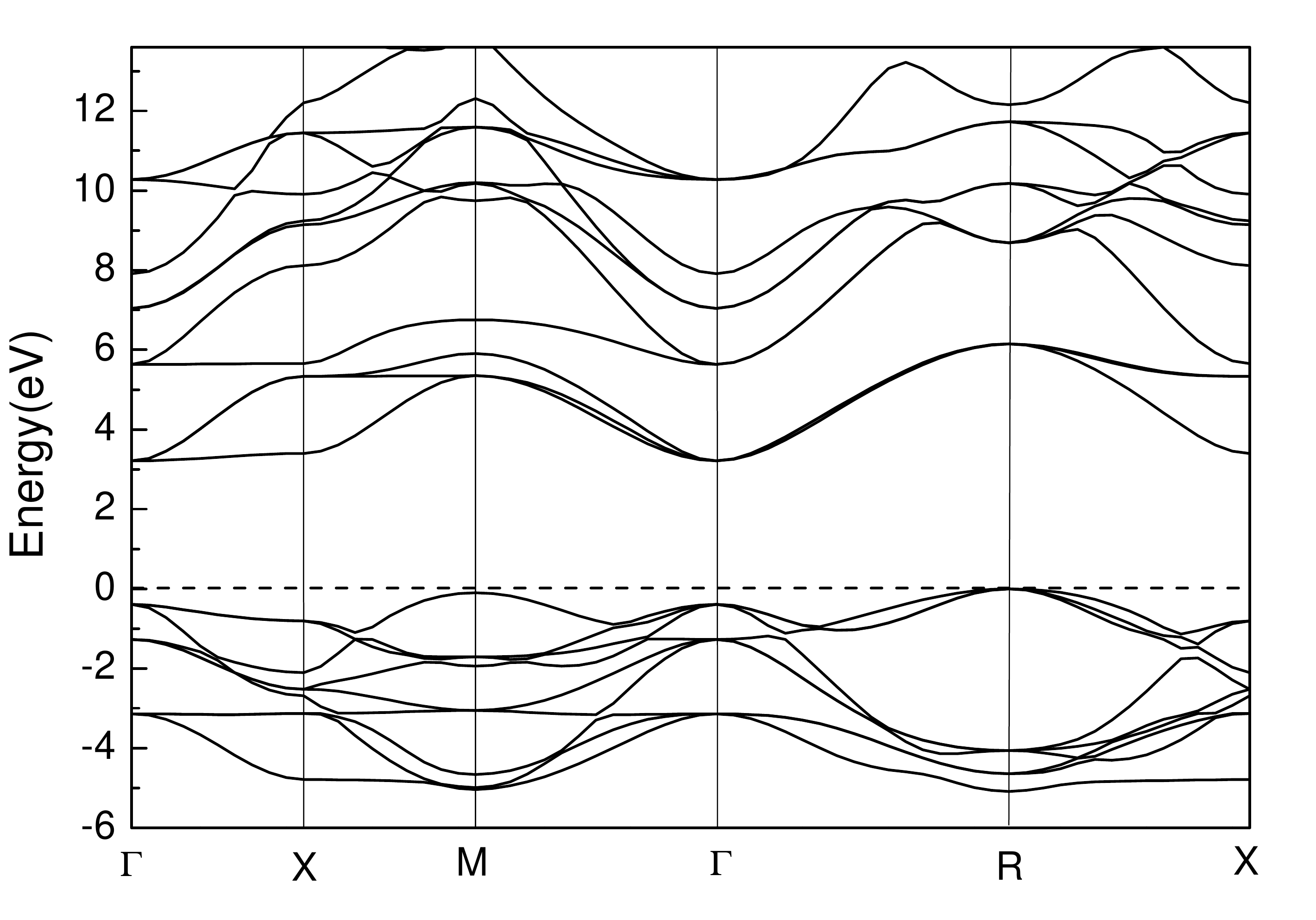} 
       \caption{Band structure of cubic  \STO~ calculated with HSE06/{\bf TZVP}. The dashed line depict the Fermi level lying at the valence band maximum ($R$ special point.) }
 \label{fig:band}
\end{figure}
\begin{figure}
     \includegraphics[width=3.0in, angle=0]{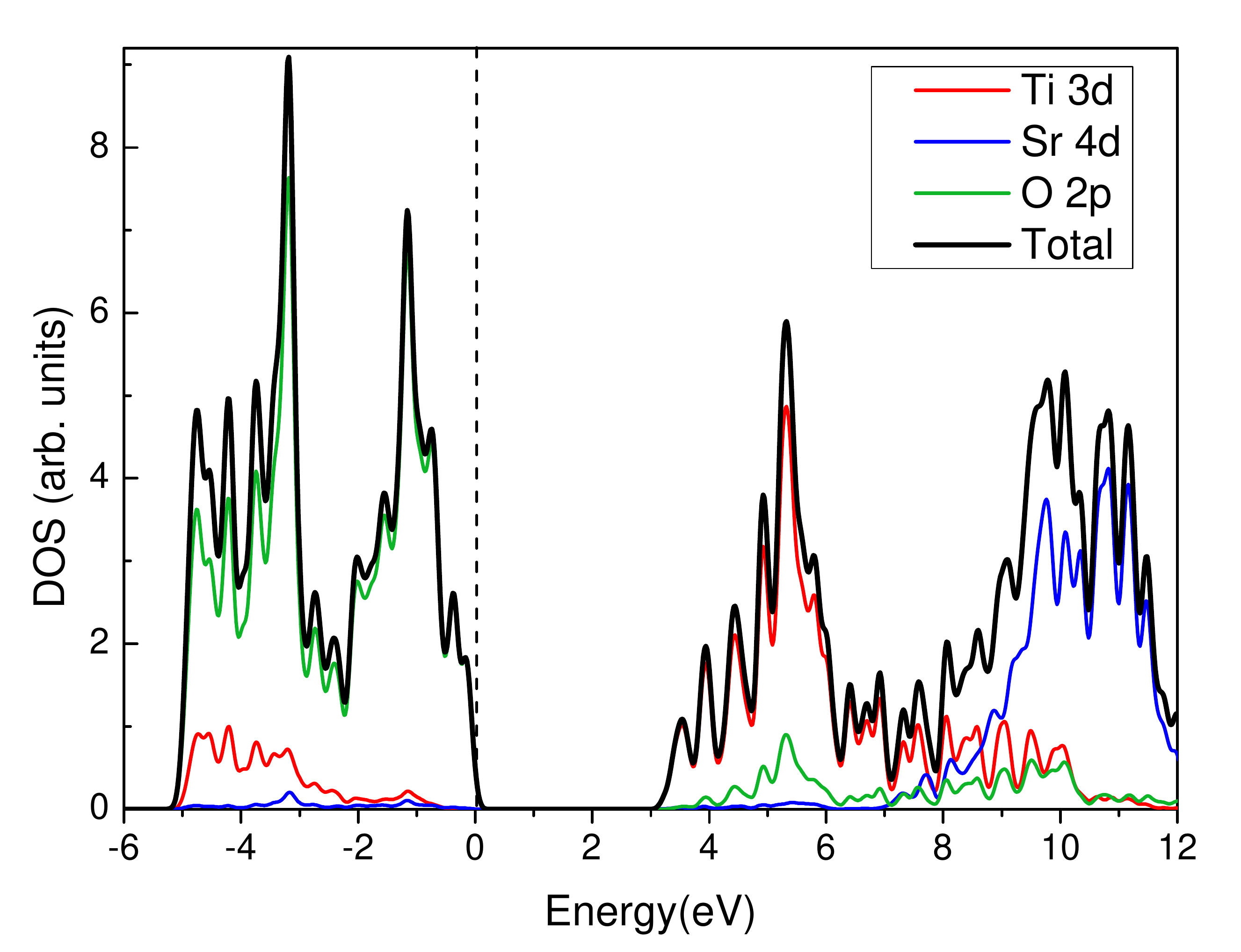}
       \caption{(Color online) Total  electronic density of states (DOS) of cubic \STO~calculated with HSE06/{\bf TZVP}. Projected density of states (PDOS) of the main contributing orbitals are also shown.}
 \label{fig:DOS}
\end{figure}
%

\subsection{Band structure alteration by the AFD phase transition}

The band structure of the cubic unit cell of STO computed with HSE06/{\bf TZVP} is shown in figure~\ref{fig:band}, with the high symmetry points $\Gamma=(0,0,0)$, $X=(0,\frac{1}{2},0)$, $M=(\frac{1}{2},\frac{1}{2},0)$ and $R=(\frac{1}{2},\frac{1}{2},\frac{1}{2})$ labeled, in the first Brillouin zone of the simple cubic system.  The dashed line depict the Fermi level lying at the valence band maximum at the $R$ point.  

Our  band structure agrees qualitatively with previous  band structures from LSDA/PW calculations,  which can be seen (for example) in Fig. 5  of Ref.~\onlinecite{Uchida:2003}, as well as the B3PW/\pisk~band structure in Ref.~\onlinecite{Pisk04} Fig. 2(a), with the exception of a few details. Our direct band gap ($\Gamma\rightarrow\Gamma$) of 3.59~eV and  indirect gap ($R\rightarrow\Gamma$) of 3.2~eV are in better agreement with experiment~\cite{Benthem:2001} compared to the underestimation observed in the  LSDA/PW gaps  and the overestimation found with B3PW/\pisk. Thus for a DFT approach, this diagram is the best band structure to date.  (An even more accurate band structure  was computed using the  experimental lattice constant of \STO~by mean of  post-LSDA quasiparticle self-consistent GW (QSGW) correction to the band structure by   Hamann and Vanderbilt.\cite{Hamann:2009})

Figure~\ref{fig:DOS} shows the total density of states (DOS) as well as the  projected density of states (PDOS) on  every atomic orbital.  The PDOS of oxygen represents the sum of the contributions of all three oxygen atoms in the cubic unit cell. In the  energy window shown here, the DOS is dominated by oxygen 2$p$, titanium 3$d$ and strontium 4$d$ states.  (All the remaining orbitals have a negligible contribution, so their PDOS are not shown.)
The valence band (VB) from 0 to $-$6~eV  is dominated by oxygen 2$p$ states, with a small contribution from titanium 3$d$ states in the range $-$3 to $-$6 eV. The conduction band (CB) is clearly dominated by titanium 3$d$ in the energy range 3.2$-$7~eV, with a smaller contribution coming from the 3 oxygen 2$p$ states as well.  
The admixture in the VB and CB between the titanium 3$d$ and oxygen 2$p$ orbitals demonstrates that the  Ti$-$O bonds have a partially  covalent character with a small degree of hybridization. (This behavior has been noted in previously published data.~\cite{Uchida:2003})
Between 7$-$9~eV, the spectrum is the sum of contributions from  oxygen 2$p$, titanium 3$d$ and strontium 4$d$ orbitals. The higher energy region in the CB (9$-$12 eV) is dominated by strontium 4$d$ orbitals with small contributions from titanium 3$d$ and with oxygen 2$p$ vanishing at around 10.5~eV. 

\begin{figure}
        \includegraphics[width=3in, angle=0]{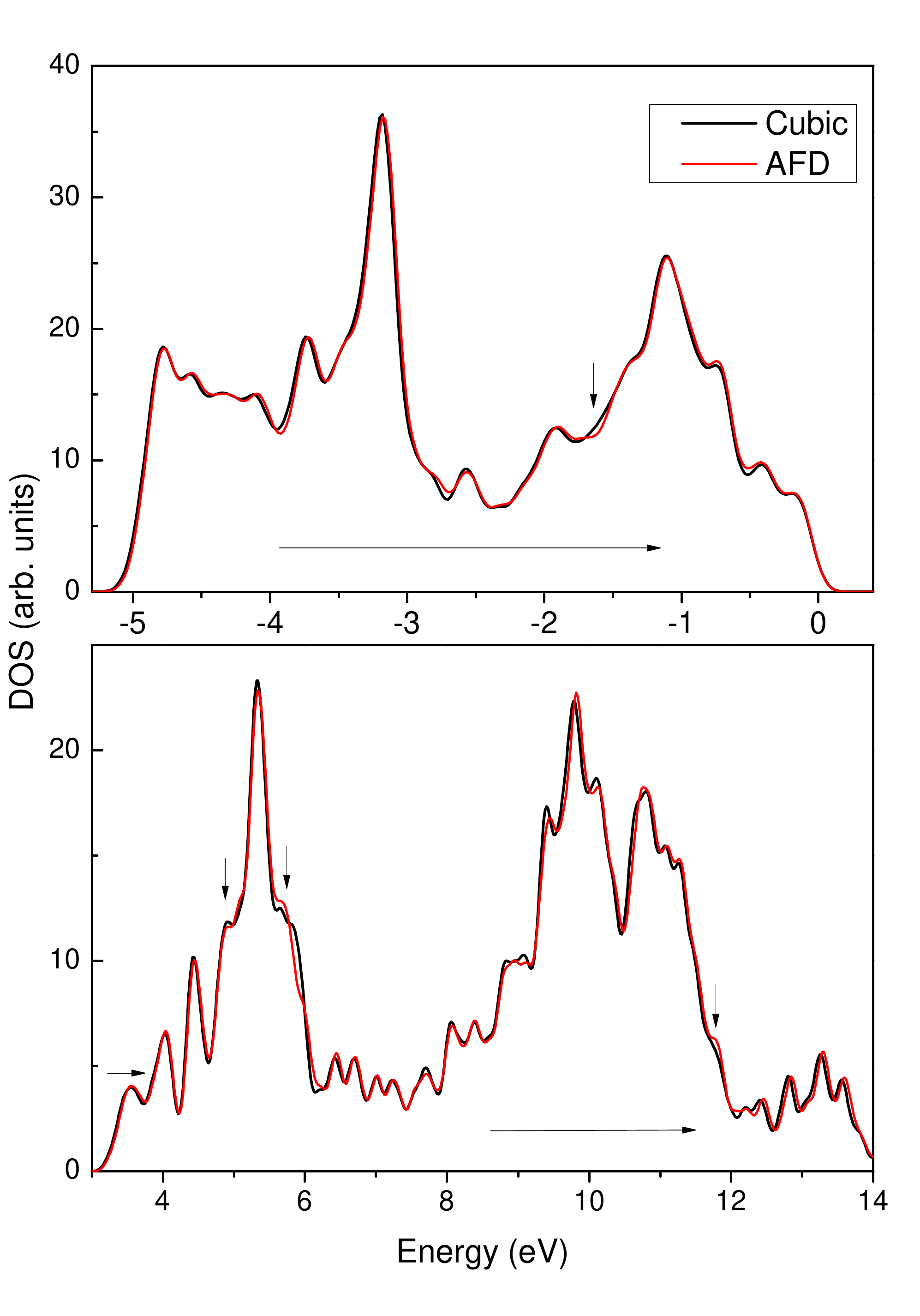}
           \caption{(Color online) Modification in the total DOS of STO upon the cubic to AFD phase transition. Horizontal arrows indicate the direction of the energy shift, while vertical arrows point to the most important changes. }
 \label{fig:CVsAFD}
\end{figure}

Figure~\ref{fig:CVsAFD} compares  the total electronic densities of states for the cubic and AFD supercells. As a general trend, the cubic to AFD phase transition does not lead to a significant modification in the total  DOS; both the valence and the conduction bands experience a slight shift to higher energies (horizontal arrows) together with some small modifications, indicated by vertical black arrows in Figure~\ref{fig:CVsAFD}. However, the VB  shift  does not affect the peak at the  VB maximum, while  a very small shift to higher energies is observed for the CB minimum, indicating that  the  band gap increase by $\approx$27 meV after the transition. 
 This same behavior holds for all the functionals/basis sets combinations tested, and is in line with some experimental observations~\cite{Hasegawa:2000} reporting very  small  changes in their measured band gaps due to the  cubic-to-tetragonal structural transition.   Further confirmation of this physical effect can be seen in recent photoluminescence  measurements,~\cite{Yamada:2010} which reported that the band gap increased by 50~meV when temperature decreased from 85~K to 8~K, which is a temperature range over which the AFD rotation would go from incomplete to nearly total.  

\begin{figure*}
        \includegraphics[width=6in, angle=0]{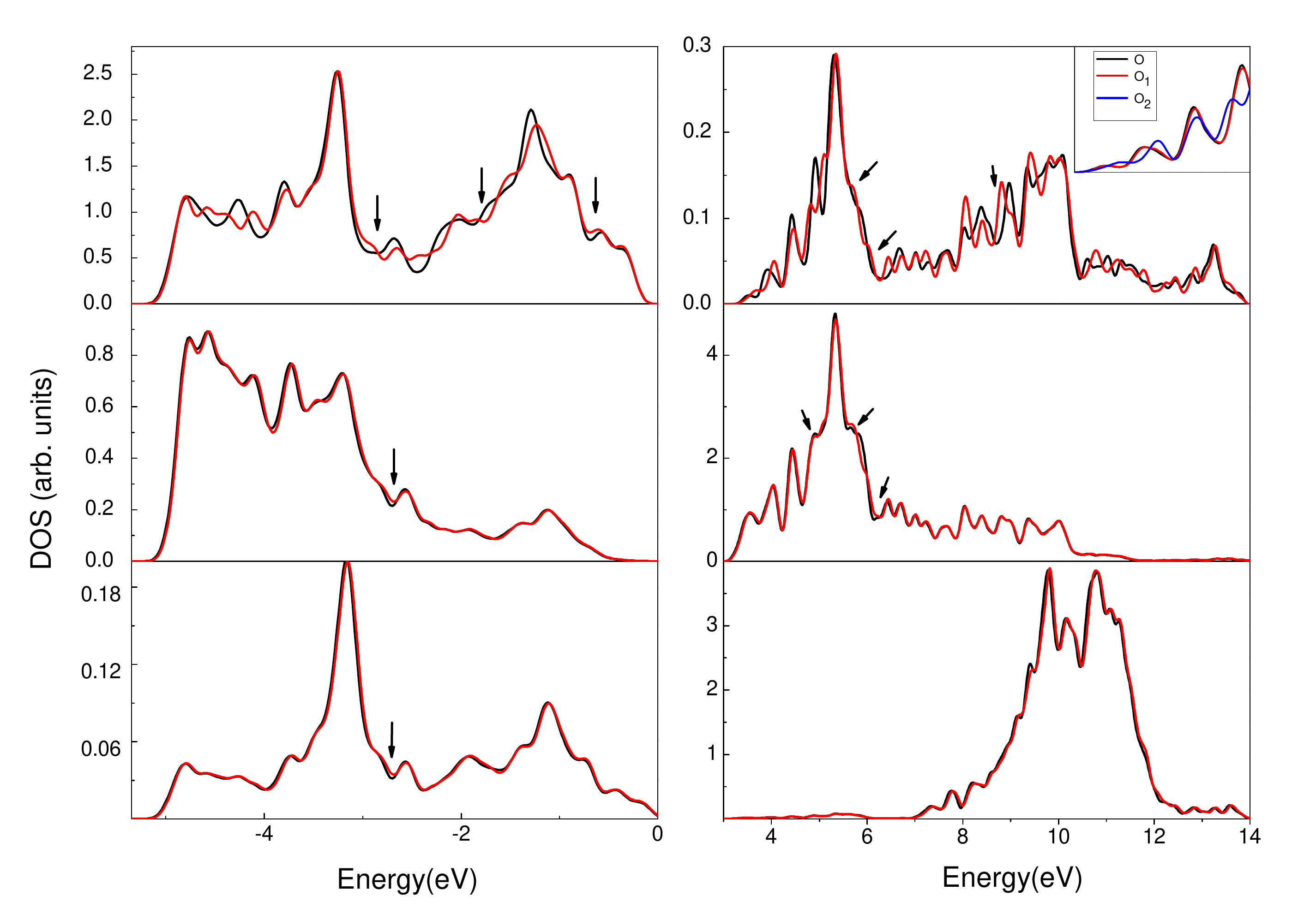}
           \caption{(Color online) Modification of the partial densities of states (PDOS) for O 2p, Ti 3d and Sr 4d  Left: Valence band Right: Conduction band}
		\label{fig:modif}
\end{figure*}

A more detailed comparison between the PDOS for each atomic orbital can give a better understanding of the origin of these modifications. It is important to mention that in the AFD supercell,  there is one nonrotating  O$_1$  atom and two rotating  O$_2$ oxygens for every Sr and Ti atom. 
Concentrating on the oxygen 2$p$ orbitals, we observe that the non-rotating  O$_1$ atoms are nearly unchanged in the PDOS compared to the cubic phase, with the exception of a tiny shift to higher energy (see the inset in Figure~\ref{fig:modif}), which can be attributed to the elongation  of the cell along the z-axis. However,  the O$_2$ demonstrate a much more significant shift to higher energies, along with changes in the height and width of some peaks.  This is  mainly  caused by the octahedral rotation involving O$_2$  atoms. The titanium 3$d$ and strontium 4$d$ spectra experience the same aforementioned shift to higher energies in the VB  and the CB due to the elongation of the lattice, with a few noticeable changes in the titanium 3$d$  spectrum  at $-$2.9 as well as between 5 and 6.5~eV.  Most of the modifications observed in the total DOS, with the exception of few originate  from the changes in the O$_2$  2$p$ and Ti 3$d$ spectra with the O$_2$ being far more important. 

\subsection {The effect of the HSE screening parameter, $\omega$} \label{sec:errors}

Relying on the assumption that plane-waves are much closer to the infinite basis
set limit than the Gaussian basis sets we used, it is useful to compare our
HSE06/{\bf TZVP} results with the HSE plane-wave results. To our knowledge, only Wahl
et al.~\cite{Wahl08} have published data  using plane-waves and  a
Heyd-Scuseria-Ernzerhof~\cite{Heyd:2003, Heyd:2004, Krukau:2006} style screened
Coulomb hybrid density functional for this system.  However,  a direct comparison with our
present  data is not possible because Wahl {\it et al.} used a different screening parameter in their calculations. 

Briefly, the HSE functional partitions the coulomb potential into short-range (SR) and long-range (LR) components:
\begin{equation}
E^{HSE}_{xc}= \frac{1}{4}E^{HF, SR}_{x}(\omega) + \frac{3}{4}E^{PBE, SR}_{x}(\omega) 
+ E^{PBE, LR}_{x}(\omega) + E^{PBE}_{c}
\end{equation}

The screening parameter $\omega$ defines the  separation range, as it controls
the distance at which the long-range nonlocal interaction becomes negligible, {\it i.e.} it ``turns off'' exact exchange after a specified distance. Wahl {\it et al.} used $\omega_1$=0.159~$a.u.^{-1}$, effectively using an
HSE-style functional, but not either of the functionals HSE03 or HSE06. Krukau {\it et al.}\cite{Krukau:2006} applied HSE while varying 0.11$\le
\omega \le$ 0.20 to a number of  bulk metals and semiconductors. They  concluded
that a small increase of $\omega$ substantially lowers the calculated band gaps
and the smaller value  omega takes in this range, the closer the calculated band gaps and lattice
constants are to the experiment. Based on the above, Krukau {\it et
al.},~\cite{Krukau:2006}  recommended $\omega_2$=0.11~$a.u.^{-1}$ for both the
HF and PBE part of the exchange. This is the  value we used in all our
calculations, and this value is part of the formal definition of HSE06. However,
 in order to make a comparison between our
HSE($\omega_2$)/TZVP and the HSE($\omega_1$)/PW data of Wahl et al., we must
perform  a  HSE/TZVP calculation  with $\omega_1$ and isolate the screening
parameter effect on the calculated properties of \STO. 

\begin{table}[!htb]
\caption{Variation of the cubic STO lattice parameter(a$_0$ in \Ang~), bulk modulus (B in GPa), and direct  ($E_g^d$) and indirect  ($E_g^i$) band gaps (in eV)  by decreasing  the HSE screening parameter from $\omega_1$=0.159 $a.u.^{-1}$ to $\omega_2$=0.11 $a.u.^{-1}$. Results are from our Gaussian basis set ({\bf TZVP}) and  the plane-wave (PW)  calculations in Ref.~\onlinecite{Wahl08}.}
\begin{ruledtabular}
\begin{tabular}{llllll}
\label{tab:omega}
	&\multicolumn{2}{c}{Gaussian} &\multicolumn{2}{c}{PW}	& Experiment\\
\cline{2-3} \cline{4-5}
\\
		& \multicolumn{2}{c}{$\omega_1\longrightarrow\omega_2$} & \multicolumn{2}{c}{$\omega_1\longrightarrow\omega_2$}\\
\\
\hline
\\
a$_0$		&3.903	&3.902	&3.904	&3.903\footnotemark[5] 	&3.890\footnotemark[1], 3.900\footnotemark[2]\\
\\
B 		&192		&193		&192		&193\footnotemark[5] 			&179\footnotemark[1], 179$\pm{4.6}$\footnotemark[3]\\
\\
$E_g^d$ 	&3.37 	& 3.59	&3.47		&3.67\footnotemark[5] 		&3.75\footnotemark[4]\\
\\
$E_g^i$ 	&2.96		&3.20		&3.07		&3.27\footnotemark[5]			 &3.25\footnotemark[4]\\
\end{tabular}
\end{ruledtabular}
\footnotetext[1]{Reference~\onlinecite{Abra95}.}
\footnotetext[2]{Reference~\onlinecite{Hell69}.}
\footnotetext[3]{Reference~\onlinecite{Fischer:1993sd}.}
\footnotetext[4]{Reference~\onlinecite{Benthem:2001}.}
\footnotetext[5]{Estimated  values if   $\omega_2$=0.11 $a.u.^{-1}$ was used in plane-wave calculations  of Ref.~\onlinecite{Wahl08}.  }
\end{table}

Table~\ref{tab:omega} shows that the HSE($\omega_1$)/TZVP  lattice constant and bulk modulus  changes very slightly    by decreasing the screening parameter  from   $\omega_1$ to $\omega_2$: 
the change is 0.001 \AA~ and  1 GPa respectively. 
A much more significant effect is, however, observed for the band gaps: decreasing the screening parameter by 50\% ( $\omega_1 \rightarrow \omega_2$), lead to an increase in the  band gaps, effectively  a rigid shift of  0.22 and 0.24~eV for the direct and indirect band gaps respectively.
If examined from the other direction, decreasing the screening parameter from $\omega_1$ to $\omega_2$ (with HSE/TZVP)  tends  to bring the band gaps  closer to the  experiment (see table~\ref{tab:omega}), which suggests that $\omega_2$ provides better agreement with experiment than $\omega_1$ does.
  The same structural changes  and band gap shifts were also found for the  small  basis sets  {\bf SZVP} and \pisk, which  are not presented here and which demonstrate that this effect is completely independent from the basis set used. 
Finally, the HSE($\omega_2$)/TZVP band gaps are very  close to the HSE($\omega_2$)/PW values we estimated,  suggesting that  our {\bf TZVP} basis set is very close in quality to
the previously used plane waves, and thus is closer to the basis set limit.

This section is contains one of the most important results of this paper, and as such 
should be clearly restated.  If we use the same version of HSE used in plane
wave studies, we can show that our {\bf TZVP} is a high quality basis set as it matches the excellent basis set plane wave results.  If
we use the proper $\omega$ in HSE with our basis set, we arrive at the best
results/smallest errors versus experiment ever reported for \STO.

Finally, it should be noted that this is not an {\it ad hoc} parameterization of $\omega$ to give the best results for this study.  We were able to obtain results that closely match experiment by using a demonstrably high quality basis set and a parameter in the density functional determined by a large test bed of structures and properties.\cite{Krukau:2006}



%
\subsection{Screened hybrids compared to regular hybrids.}\label{hybrids}
%
\begin{table}[!htb]

\caption{Our most converged direct  ($E_g^d$) and indirect  ($E_g^i$) band gaps (in eV) for cubic STO alongside  previously published hybrid  functional results done with the \pisk~basis set.  Regular hybrids data are corrected according to the  basis set sensitivity effect deduced in section~\ref{sec:elect_cub}. }
\begin{ruledtabular}
\begin{tabular}{llllll}
\label{tab:all_litt}

 &functional/basis &\multicolumn{2}{c}{$E_g^d$} &\multicolumn{2}{c}{$E_g^i$ }\\
\cline{3-4} \cline{5-6} 
\\
 & &Ori. &Corr.$^\text{estimated}$ &Ori. &Corr.$^\text{estimated}$\\
\hline
\\
Exp.~\cite{Benthem:2001}                &   &3.75 &&3.25\\
                                    
Present                            &HSE06/{\bf TZVP} &3.59 &&3.20\\
Ref.~\onlinecite{Pisk04}               &B3PW/\pisk &3.96 &3.74   &3.63 &3.35\\
					&B3LYP/\pisk &3.89 &3.67    &3.57 &3.30\\
Ref.~\onlinecite{Heif06}                  &B3PW/\pisk  &4.02 &3.80   &3.70 &3.42\\
Ref.~\onlinecite{Zhuk09}           &B3PW/\pisk &---&---      &3.63 &3.35\\
Ref.~\onlinecite{Bilc08}   &B1-WC/\pisk\footnotemark[1]   &3.91   & &3.57\\
\\

\end{tabular}
\end{ruledtabular}
\footnotetext[1]{\pisk~basis set with all electrons for Ti, basis set correction cannot be applied.}
\end{table}
Table~\ref{tab:all_litt} summarizes the calculated band gaps of HSE06/{\bf TZVP} and compares them with previously published  gaps computed with the regular hybrids B3PW and B3LYP, done with the  \pisk~basis set.  There are noticeable  differences between the results of HSE06 and the regular hybrids, with HSE06/{\bf TZVP} giving  band gaps very close to experiment while  regular hybrids used with \pisk~ overestimate the gap, especially the indirect band gap. The band gap overestimation is of same  magnitude we observed in  section~\ref{sec:elect_cub} for HSE06/\pisk~as well as all the other functionals tested on STO with \pisk. This suggests that \pisk~is also behind the band gap overestimation  in the  regular hybrids data reported in the literature.\cite{Pisk04,Heif06,Zhuk09}  By comparing the \pisk~ and {\bf TZVP} band gaps from table~\ref{tab:elec_cub}, we can  deduce that  the \pisk~ basis set has an effect (versus a large basis set) of increasing  the direct and indirect band gaps by average values of  0.22 and 0.28~eV  respectively.  By applying  this  \pisk$\rightarrow${\bf TZVP}  basis set correction to the regular hybrid  B3PW/\pisk~and B3LYP/\pisk~ band gaps  (see the corrected values in table~\ref{tab:all_litt}), band gaps are brought closer to the experimental values, and thus closer to the HSE06/{\bf TZVP} results as well. Consequently, differences in the computed electronic properties of HSE06 and B3PW and B3LYP  are considerably attenuated  and suggest that the screened hybrid  HSE06 is  comparable in accuracy with regular hybrids for STO, while being much more computationally efficient.

 The final issue to discuss is the comparison of the structural and elastic properties of STO computed with HSE06 versus regular hybrids.
 Perovskite crystals in the  cubic structure have only three independent elastic constants, namely C$_{11}$, C$_{12}$ and C$_{44}$, as well as a bulk modulus:
\beeq
\label{eq:bulk}
B=\frac{1}{3}(C_{11}+2C_{12})
\eneq  
We calculated the elastic constants of STO using HSE06/TVZP,  following the methodology described in Ref.~\onlinecite{Wu:2005}. Ideally we would like to compare our cubic elastic constants  calculated at 0~K with low temperature data, but as experimentally the cubic structure turns to a tetragonal structure below a transition temperature, making any comparison of this kind impossible.  
Experimentally, Bell and Rupprecht~\cite{Bell:1963} found  that the elastic constants of STO measured between  303 and 112~K obey the following empirical relations: 

\begin{subequations}
\beeqa
\label{eq:ela}
C_{11}=334.1[1-2.62\times10^{-4}(T-T_a) -\frac{0.0992}{(T-T_a)}]\\
C_{12}=104.9[1-1.23\times10^{-4}(T-T_a) +\frac{0.1064}{(T-T_a)}]\\
C_{44}=126.7[1-1.30\times10^{-4}(T-T_a) -\frac{0.1242}{(T-T_a)}]
\eneqa
\end{subequations}

\noindent where  the elastic constants are in GPa,  T is the temperature and  $T_a$=108~K is the critical temperature. C$_{11}$ and C$_{44}$ reach their maximum values at 133~K where STO is still cubic, then  they start to decrease as $-1/(T-T_a)$ in the region around the transition temperature, in contrast,  C$_{12}$ continue to  increase as $1/(T-T_a)$ in the same temperature range.

 Since we don't know at which temperature the change from cubic to tetragonal phase begins to take place, it is better to limit our comparison with data measured at 133~K and above. 
Table~\ref{tab:ela} summarizes  our results and compares them with  experiment as well as  previously published results with B3PW/\pisk~ and B3LYP/\pisk.  HSE06/{\bf TZVP} provides excellent lattice constants but  predicts the bulk modulus to be  8\% higher than experiment. The elastic constants from HSE06/{\bf TZVP} overestimate the  experimental data at room temperature  by 10\% and  the 133~K data by 6\%; this was expected given the overestimation of the bulk modulus. The B3PW hybrid also  gave very good lattice constant and bulk modulus,  but the calculated elastic constants are  lower than  the room and the low temperature experimental values.  B3LYP predicted a lattice constant higher  by 1\%, a good bulk modulus, and offers the best agreement with the low temperatures elastic constants. 
In summary, none of the screened or regular hybrids considered was able to give simultaneously excellent  bulk moduli and elastic constants, still  HSE06/{\bf TZVP}  offer the best compromise between  efficiency, accuracy and speed.  
\begin{table}[!htb]

\caption{Calculated elastic constants with HSE06/{\bf TZVP} for cubic STO compared to experiment and  previously published results with the regular hybrid  functional B3PW and the \pisk~basis set from Ref.~\onlinecite{Pisk04}. a$_0$ is in \Ang, B, C$_{11}$, C$_{12}$ and C$_{44}$ are in GPa}
\begin{ruledtabular}
\begin{tabular}{llllll}
\label{tab:ela}

\\
 &$a_0$    &B &C$_{11}$ &C$_{12}$ &C$_{44}$ \\
\hline
\\
HSE06/{\bf TZVP}                          &3.902    & 193  &351.4  &113 &137.3 \\

B3PW/\pisk               &3.900   &177    &316 &92.7& 120.1\\
B3LYP/\pisk                &3.940   &177    &328.3 &105.7 & 124.6\\
Exp.                    &3.890~\cite{Abra95}     &179~\cite{Abra95}    &317.2 &102.5 &123.5\footnotemark[1]  \\
                                     &3.900~\cite{Hell69} &179$\pm{4.6}$~\cite{Fischer:1993sd} &330 &105 &126\footnotemark[2]\\
 &3.910 &184 & & &128\footnotemark[3]
\\

\end{tabular}
\end{ruledtabular}
\footnotetext[1]{Ref. ~\onlinecite{Bell:1963} at room temperature.}
\footnotetext[2]{Ref. ~\onlinecite{Bell:1963}: max. measured values for C$_{11}$ and  C$_{44}$  at 133~K, C$_{12}$  increase further as temperature drop.}
\footnotetext[3]{Landolt–B\"ornstein Group III Condensed Matter 2002 vol 36, subvol V (Berlin: Springer) chapter 1A (Simple Perovskyte-Type Oxides) pp 116–47}
\end{table}

\section{Conclusion}\label{sec:conc} 

We used the {\it ab-initio} code {\sc{gaussian}} to simulate the properties of \STO~(STO) using a large spectrum of functionals, from LSDA, GGAs (PBE and PBEsol) and meta-GGAs (M06L and revTPSS) to modern range-separated hybrid functionals (HSE06 and HISS); assessing their  ability in  predicting  the properties of the cubic and the AFD phases of STO.  

We found that pure DFT functionals tend to overestimate the octahedral rotation angles  of the AFD phase, in agreement with previously reported results in the literature using  plane-wave basis sets of comparable quality.~\cite{Wahl08} Also, basis sets of low quality tend to inhibit  the tetragonality of the AFD phase and sometimes even suppress it, regardless of the functional used.   We therefore constructed a localized basis set of sufficient completeness (or size) to correctly simulate the \tio octahedral rotation  and the cubic phases of STO.  We also evaluated  the band gap errors arising from the use \pisk~basis set  and from the magnitude of the HSE screening parameter $\omega$. 
By applying our  basis set and $\omega$ corrections to the previously published work with regular and screened hybrid functionals  on STO,  we showed that the discrepancies between published simulated data can be explained and that hybrid functionals used with sufficiently big Gaussian-type basis sets can  give results comparable with plane-wave calculations and in excellent agreement with experiment.  

The screened hybrid functional HSE06 predicts  the electronic and structural properties of the cubic and  AFD phase in very good agreement with experiment, especially if used with high quality basis set {\bf TZVP}. 
HSE06/{\bf TZVP} is the  most reliable combination of functional and Gaussian basis set for STO which is  computationally tractable with the current computer power.  It is  accurate enough to  enable us to  understand  the changes in the band structure during the cubic to AFD phase transition.  The success of  HSE06/{\bf TZVP} encourages its use on more complicated cases like the bond breaking and over binding and defect formation, where the basis set completeness is expected to play a major role. 

\begin{acknowledgments}

This work is supported by the  Qatar National Research Fund  (QNRF) through the National Priorities  Research Program (NPRP 481007-20000). We are thankful to Cris V. Diaconu for the technical support with the band structure code in {\sc{gaussian}}. We are grateful to the Research computing facilities at Texas A\&M university at Qatar for generous allocations of computer resources.\\

\end{acknowledgments}

\end{document}